\documentclass[showpacs,preprintnumbers]{revtex4}
\usepackage{graphicx}
\usepackage{amsmath,amssymb}
\usepackage{subfigure}

\def\bc{\begin{center}}
\def\ec{\end{center}}

\def\beq{\begin{equation}}
\def\eeq{\end{equation}}

\begin{document}

\title{Superfluidity and collective properties of excitonic polaritons in gapped graphene in a
microcavity}

\author{Oleg L. Berman$^{1,2}$, Roman Ya. Kezerashvili$^{1,2}$, and Klaus
Ziegler$^{1,3}$}
\affiliation{\mbox{$^{1}$Physics Department, New York City College
of Technology, The City University of New York,} \\
Brooklyn, NY 11201, USA \\
\mbox{$^{2}$The Graduate School and University Center, The
City University of New York,} \\
New York, NY 10016, USA \\
\mbox{$^3$   Institut f\"ur Physik, Universit\"at Augsburg\\
D-86135 Augsburg, Germany }}
\date{\today}

\begin{abstract}

We predict the formation and superfluidity of polaritons in an optical
microcavity formed by excitons in gapped graphene embedded there and microcavity photons. The Rabi
splitting related to the creation of an exciton in a graphene layer in the presence of the band
gap is obtained.  The analysis of collective excitations as well as the sound velocity is
presented. We show that the superfluid density $n_{s}$ and temperature of
the Kosterlitz-Thouless phase transition $T_{c}$ are decreasing functions of
the energy gap.

\end{abstract}

\maketitle


\section{Introduction}

\label{intro}

Up today all theoretical and experimental studies have been devoted to Bose
coherent effects of two-dimensional (2D) excitonic polaritons in a quantum
well embedded in a semiconductor microcavity~\cite%
{book,pssb,Littlewood,Snoke_text}. To obtain polaritons, two mirrors
placed opposite each other in order to form a microcavity, and
quantum wells are embedded within the cavity at the antinodes of the
confined optical mode. The resonant exciton-photon interaction
results in the Rabi splitting of the excitation spectrum. Two
polariton branches appear in the spectrum due to the resonant
exciton-photon coupling. The lower polariton branch of the spectrum
has a minimum at zero momentum. The effective mass of the lower
polariton is extremely small. These lower polaritons form a 2D
weakly interacting Bose gas. The extremely light mass of these
bosonic quasiparticles at experimentally achievable excitonic
densities, results in a relatively high critical temperature for
superfluidity, because the 2D thermal de Broglie wavelength,  which
becomes comparable to the distance between the bosons, is inversely
proportional to the mass of the quasiparticle.

Recently there were many experimental and
theoretical studies devoted to graphene known by unusual properties in its
band structure~\cite{Castro_Neto_rmp,Das_Sarma_rmp}.  Due to
the absence of a gap between the conduction and valence bands in graphene,
the screening effects result in the absence of electronic excitations in
graphene. Today we achieved different ways
to obtain a gap in graphene. For example, the gap in graphene
structures can be formed due to magnetic field, doping, electric
field in biased graphene and finite size quantization in graphene
nanoribbons. A gap in the electron spectrum in graphene can be
opened by applying the magnetic field, which results in the
formation of magnetoexcitons~\cite{Iyengar}. Excitons in graphene
can be also formed due to gap opening in the electron and hole
spectra in the graphene layer by doping \cite{levitov10}. There were
a number of papers devoted to the excitonic effects in different
graphene-based structures. Significant excitonic effects related to
strong-electron-hole correlations were observed in graphene by
measuring its optical conductivity in a broad spectral
range~\cite{Heinz}. The observed excitonic resonance was explained
within a phenomenological model as a Fano
interference of a strongly coupled excitonic state and a band continuum~\cite%
{Heinz}. The electron-hole pair condensation in the two graphene
layers have been studied in
Refs.~\onlinecite{Sokolik,Joglekar,MacDonald1,MacDonald2,Efetov}.
The possibility of formation of edge-state excitons in graphene
nanoribbons is caused by the appearance of the gap in the electron
energy spectrum due to the finite size quantization. A
first-principles calculation of the optical properties of
armchair-edged graphene nanoribbons with many-electron and excitonic
effects included was presented in Ref.~\onlinecite{Louie_nl_2007}.
In Ref.~\onlinecite{Louie_PRL_2008} were studied the optical
properties of zigzag-edged graphene nanoribbons  with the spin
interaction. It was found that optical response was dominated by
magnetic edge-state-derived excitons with large binding energy.
First-principles calculations of many-electron effects on the
optical response of graphene, bilayer graphene, and graphite were
described in Ref.~\onlinecite{Louie_PRL_2009}. It was found that
resonant excitons were formed in these two-dimensional semimetals.
The other mechanism of electronic excitations in graphene can be
achieved in biased graphene, where the energy band gap is formed by
applied electric field. A continuously tunable bandgap of up to
$250\ \mathrm{meV}$ was generated in biased bilayer
graphene~\cite{Zhang}. It was shown that the
optical response of this system is dominated by bound excitons~\cite%
{Louie_nl_2010}.

According to Ref.~\onlinecite{Haberer} a tunable gap in graphene can
be induced and controlled by hydrogenation. The excitons in gapped
graphene can be created by laser pumping. The superfluidity of
quasi-two-dimensional dipole excitons in double-layer graphene in
the presence of band gaps was proposed recently in
Ref.~\onlinecite{BKZ}.

The Bose-Einstein condensation (BEC) of magnetoexcitonic polaritons
formed by magnetoexcitons in graphene embedded in a semiconductor
microcavity in a high magnetic field in a planar harmonic potential
trap was studied in Ref.~\onlinecite{BKL}. However, the interaction
between two direct 2D magnetoexcitons in graphene is negligibly
small in a strong magnetic field, in analogy to 2D magnetoexcitons
in a quantum well~\cite{Lerner}. Therefore, the superfluidity of
magnetoexcitonic polaritons in graphene in this case is absent,
since the superfluidity is caused the sound spectrum of Bose
collective excitations due to the exciton-exciton interaction which
is negligible in graphene in a high magnetic field.

In this paper we consider
the direct 2D excitons formed in a single graphene layer in the presence of
the band gap and predict the superfluidity of polaritons formed by these
excitons and microcavity photons, when the graphene layer is embedded into
an optical microcavity We obtained the corresponding superfluid density and
temperature for the Kosterlitz-Thouless phase transition due to the
superfluidity of microcavity polaritons.

The paper is organized in the following way. In Sec.~\ref{Hami} we
presented the Hamiltonian of excitons in a graphene layer embedded
in an optical microcavity. In Sec.~\ref{ex-ham} we obtain the
excitonic Hamiltonian which is the sum of the Hamiltonian of
non-interacting excitons in the gapped graphene and Hamiltonian that
describes the exciton-exciton interaction. The
Hamiltonian of photons in a semiconductor microcavity is given in Sec. \ref%
{Micro photons}. In Sec. \ref{ex-ph} the Hamiltonian of the harmonic
exciton-photon coupling in the gapped graphene is derived and
corresponding Rabi splitting constant is obtained. The study the
condensation of a gas of microcavity polaritons, the density of the
superfluid component, as well as the Kosterlitz-Thouless temperature
are presented in Sec.~\ref{sup}.
Finally, the discussion of the results and the conclusions follow in Sec. %
\ref{disc}.

\section{Hamiltonian of gapped graphene excitons in microcavity}

\label{Hami}

The total Hamiltonian $\hat{H}_{tot}$ of the system of 2D excitons in gapped
graphene embedded in an optical microcavity and 2D microcavity photons can
be written as
\begin{equation}
\hat{H}_{tot}=\hat{H}_{ex}+\hat{H}_{ph}+\hat{H}_{ex-ph}\ ,
\label{tot_ham}
\end{equation}
where $\hat{H}_{ex}$ is the Hamiltonian of excitons in graphene in
the presence of the gap, $\hat{H}_{ph}$ is a Hamiltonian of photons
in a microcavity, and $\hat{H}_{ex-ph}$  is a Hamiltonian of
exciton-photon interaction.

The Hamiltonian of 2D excitons in the graphene in the presence of a gap is
given by
\begin{equation}
\hat{H}_{ex}=\hat{H}_{ex}^{(0)}+\hat{H}_{ex-ex},  \label{Ham_exc}
\end{equation}%
where $\hat{H}_{ex}^{(0)}$ is the Hamiltonian of non-interacting 2D excitons
in gapped graphene and $\hat{H}_{ex-ex}$ is the Hamiltonian of the
exciton-exciton interaction.

The Hamiltonian of non-interacting excitons in gapped graphene $\hat{H}%
_{ex}^{(0)}$ is given by
\begin{equation}
\hat{H}_{ex}^{(0)}=\sum_{\mathbf{P}}^{{}}\epsilon _{ex}(P)\hat{b}_{%
\mathbf{P}}^{\dagger }\hat{b}_{\mathbf{P}}^{{}}\ ,  \label{Ham_exc0}
\end{equation}%
where $\hat{b}_{\mathbf{P}}^{\dagger }$ and $\hat{b}_{\mathbf{P}}$ are
excitonic creation and annihilation operators obeying to Bose commutation
relations and $\epsilon _{ex}(P)$ is the energy dispersion of a single
exciton in a graphene layer.

The Hamiltonian  of the exciton-exciton interaction
$\hat{H}_{ex-ex}$  in graphene in the presence of a gap is given by
\begin{equation}
\hat{H}_{ex-ex}=\frac{1}{2A}\sum_{\mathbf{P},\mathbf{P}^{\prime },\mathbf{q}%
}U_{\mathbf{q}}\hat{b}_{\mathbf{P}+\mathbf{q}}^{\dagger }\hat{b}_{\mathbf{P}%
^{\prime }-\mathbf{q}}^{\dagger }\hat{b}_{\mathbf{P}}\hat{b}_{\mathbf{P}%
^{\prime }},  \label{Ham_exc_int}
\end{equation}
where $A$ is the macroscopic quantization area and $U_{\mathbf{q}}$ is the
Fourier transform of the exciton-exciton pair repulsion potential.

The Hamiltonian of photons in a semiconductor microcavity is given by~\cite%
{Pau}:
\begin{equation}
\hat{H}_{ph}=\sum_{\mathbf{P}}\epsilon _{ph}(P)\hat{a}_{\mathbf{P}%
}^{\dagger }\hat{a}_{\mathbf{P}}^{{}},  \label{Ham_ph}
\end{equation}
where $\hat{a}_{\mathbf{P}}^{\dagger }$ and $\hat{a}_{\mathbf{P}}$
are the photonic creation and annihilation Bose operators, and
$\epsilon _{ph}(P)$ is the cavity photon energy dispersion.

Follow Ref.~\onlinecite{Ciuti} the Hamiltonian of the harmonic
exciton-photon coupling can be written as

\begin{equation}
\hat{H}_{ex-ph}={\hbar \Omega _{R}}\sum_{\mathbf{P}}\hat{a}_{\mathbf{P}%
}^{\dagger }\hat{b}_{\mathbf{P}}^{{}}+h.c.,  \label{Ham_exph}
\end{equation}
where the exciton-photon coupling energy represented by the Rabi
splitting constant $\hbar \Omega _{R} $ is obtained below.

Let us present the detailed consideration and analysis of each term in (\ref%
{tot_ham}): firstly, the excitonc Hamiltonian $\hat{H}_{ex}$ that describes
the formation excitons in gapped graphene, secondly, the Hamiltonian $\hat{H}%
_{ph}$ that describes the microcavity photons and lastly, the Hamiltonian $%
\hat{H}_{ex-ph}\ $responsible for the exciton-photon coupling within the
microcavity.

\section{Excitonic Hamiltonian}

\label{ex-ham}

In this Section we present the excitonic Hamiltonian $\hat{H}_{ex}$ in Eq.~(%
\ref{tot_ham}) that consist from two terms: the Hamiltonian that
describes the formation of the gas of non-interacting excitons in
gapped graphene and Hamiltonian responsible for the exciton-exciton
interaction that we assume is strong enough to be neglected.

\subsection{An exciton in a graphene layer in the presence of the gap}

\label{single}

The Hamiltonian of 2D excitons in the gapped graphene is given by
Eq.~(\ref{Ham_exc}). As the first step we analyze the Hamiltonian of
non-interacting excitons
$\hat{H}_{ex}^{(0)}$. As it follows from Eq.~(\ref%
{Ham_exc0}) this Hamiltonian is determined by the energy-momentum
despersion of non-interacting excitons $\epsilon _{ex}(P).$
Therefore, we need to find out the energy-momentum despersion
$\epsilon _{ex}(P)$ of the electron and hole that are bound via an
electromagnetic attraction  by solving the two-body problem in a
gapped graphene layer.

We consider an electron and a hole located in a single graphene
sheet and introduce the gap parameter $\delta$. We assume that our
exciton is formed
by the electron and hole located in one graphene sheet. The gap parameter $%
\delta$ is the consequence of adatoms on the graphene sheets (e.g, by
hydrogen, oxygen or other non-carbon atoms \cite{levitov10}) which create a
one-particle potential.

We use the coordinate vectors of the electron and the hole $\mathbf{r}_1$
and $\mathbf{r}_2$. Each honeycomb lattice is characterized by the
coordinates $(\mathbf{r}_j,1)$ on sublattice A and $(\mathbf{r}_j,2)$ on
sublattice B. Then the two-particle wave function, describing two particles
in the same graphene sheet, reads $\Psi(\mathbf{r}_1,s_1;\mathbf{r}_2,s_2)$.
This wave function can also be understood as a four-component spinor, where
the spinor components refer to the four possible values of the sublattice
indices $s_1,s_2$:
\begin{equation}  \label{wavefunction1}
\Psi(\mathbf{r}_1,s_1;\mathbf{r}_2,s_2)=\left({%
\begin{array}{c}
\phi_{aa}(\mathbf{r}_1,\mathbf{r}_2) \\
\phi_{ab}(\mathbf{r}_1,\mathbf{r}_2) \\
\phi_{ba}(\mathbf{r}_1,\mathbf{r}_2) \\
\phi_{bb}(\mathbf{r}_1,\mathbf{r}_2)%
\end{array}%
}\right) \ .
\end{equation}

The system of two interacting particles located in the same graphene
sheet by using the coordinates $\mathbf{r}$, which is the projection
of the difference vector between between an electron and a hole
$\mathbf{r}_1-\mathbf{r}_2$ on the plane parallel to the graphene
sheet. Then the Hamiltonian of the two-body problem with a broken
sublattice symmetry can be described by the Hamiltonian
\begin{equation}
\mathcal{H}=\left(
\begin{array}{cccc}
V(r) & d_{2} & d_{1} & 0 \\
d_{2}^{\dagger} & -2\delta +V(r) & 0 & d_{1} \\
d_{1}^{\dagger} & 0 & 2\delta +V(r) & d_{2} \\
0 & d_{1}^{\dagger} & d_{2}^{\dagger} & V(r)%
\end{array}%
\right) \ ,  \label{ham1}
\end{equation}%
%
where $V(r)$ is the electron-hole electromagnetic  interaction as
function of $r$, which is the projection of the vector
$\mathbf{r}_{1}-\mathbf{r}_{2}$ on the plane parallel to the
graphene sheet. In Eq.~(\ref{ham1}), $d_{j}=\hbar
v_{F}(-i\partial_{x_{j}} -
\partial_{y_{j}})$, $d_{j}^{\dagger }=\hbar
v_{F}(-i\partial_{x_{j}}+
\partial_{y_{j}})$, where $\partial_{x_{j}} =
\partial/\partial x_{j}$, $\partial_{y_{j}} =
\partial/\partial y_{j}$,
$j=1,2$, and $x$ and $y$ are components of the vectors $\mathbf{r}_{1}$ and $%
\mathbf{r}_{2}$ that represent the coordinates of the electron and
hole, correspondingly, and $v_{F}=\sqrt{3}at/(2\hbar )$ is the Fermi
velocity of electrons in graphene, where $a=2.566\
\mathrm{\mathring{A}}$ is a lattice constant and $t\approx 2.71\
\mathrm{eV}$ is the overlap integral between the nearest carbon
atoms \ \cite{Lukose}. In Eq.~(\ref{ham1}) we take into account the
renormalization of the electron-hole distance in the electron-hole
Coulomb attraction due to the non-locality of electron and hole wave
function
assuming the following model for the electron-hole attraction $%
V(r)=-e^{2}/(4\pi \epsilon _{0}\epsilon (r^{2}+r_{0}^{2})^{1/2})$,
where $r_{0}$ is the renormalization parameter which will be estimated, $%
\epsilon _{0}=8.85\times 10^{-12}\ \mathrm{C^{2}/(N m^{2})}$, $e$
is the electron charge, $\varepsilon $ is the dielectric constant of the
material surrounding graphene sheet.  It should be mentioned that the main
contribution to the polariton mass is the cavity photon mass rather
than exciton mass, and, therefore, we use our model to estimate
the exciton mass roughly by the order of magnitude. Assuming $r\ll r_{0}$,
we expand $V(r)$ in Taylor series as $V(r)=-V_{0}+\gamma r^{2}$, where $%
V_{0}=e^{2}/(4\pi \varepsilon \varepsilon _{0}r_{0})$ and $\gamma
=e^{2}/(8\pi \varepsilon _{0}\varepsilon r_{0}^{3})$.

Now we have to solve the eigenvalue problem of the Hamiltonian $\mathcal{H}%
\Psi=\epsilon\Psi$ for the energy of the exciton $\epsilon$. The
eigenfunction depends on the coordinates of both particles, namely
$(\mathbf{r}_1,\mathbf{r}_2)$. To separate the relative motion of
the electron and hole we use the following ansatz for the wave
function
\begin{equation}
\Psi_j(\mathbf{R},\mathbf{r})=\mathtt{e}^{i \mathbf{P}\cdot
\mathbf{R}/\hbar}\;\psi_j(\mathbf{r}) \ ,
\end{equation}
where  $\mathbf{P}$ is momentum, and following to
Ref.~\onlinecite{BKZ} for a generalized center of mass coordinate
$\mathbf{R}$ and relative coordinate $\mathbf{r}$ we have

\begin{gather}  \label{exp1}
\mathbf{R}=\alpha \mathbf{r}_1+\beta \mathbf{r}_2 \ , \ \ \ \ \ \ \mathbf{r}
= \mathbf{r}_1-\mathbf{r}_2 \ ,
\end{gather}
with the parameters
\begin{eqnarray}  \label{albet}
\alpha=\frac{\epsilon- 2 \delta}{2 \epsilon} \ , \ \ \ \ \ \ \ \ \ \beta=%
\frac{\epsilon+ 2\delta}{2\epsilon} \ .
\end{eqnarray}

The procedure given in~Ref.~\onlinecite{BKZ} can be applied to the
eigenvalue problem $(\mathcal{H}+V_0)\Psi=(\epsilon + V_{0})\Psi$
when we assume that both relative and center-of-mass kinetic
energies, as well as  the harmonic term in the potential energy
$V(r)+ V_0 = \gamma r^{2}$ are small in comparison to the gap energy
$2\delta$. Starting from the wave function in
Eq.~(\ref{wavefunction1}) and by using the same procedure as we
applied in Ref.~\onlinecite{BKZ} we obtain for the spinor component
$\phi_{aa}$ the equation
\begin{equation}  \label{fin11}
\left({\ \frac{(v_F P)^{2}}{2\epsilon}+V(r)-\frac{2\epsilon
(\hbar v_F)^2 \nabla_\mathbf{r}^2}{\left({\epsilon^2 - 4\delta^2}\right)}}%
\right)\phi_{aa}= \epsilon \phi_{aa} \ .
\end{equation}
Let us rewrite  Eq.~(\ref{fin11}) in more convenient form
\begin{eqnarray}
\left( {-\mathcal{F}\nabla _{r}^{2}+\gamma r^{2}}\right) \phi_{aa}=\mathcal{F}_{0}\phi _{aa}\
,
\label{harm}
\end{eqnarray}%
where $\mathcal{F}$ is given by
\begin{eqnarray}
\label{F} \mathcal{F}=\frac{2(\epsilon + V_{0}) (\hbar
v_{F})^{2}}{\left( ({\epsilon + V_{0}) ^{2}-4\delta ^{2}}\right) }\
\end{eqnarray}%
and   $\mathcal{F}_{0}$ is given by
\begin{eqnarray}
\label{F0} \mathcal{F}_{0}=\epsilon + V_{0} -\frac{%
(v_{F}P)^{2}}{2(\epsilon + V_{0})} \ .
\end{eqnarray}%
Eq.~(\ref{harm}) describes a two-dimensional isotropic harmonic
oscillator, whose solutions are given by the condition
\begin{equation}
\frac{\mathcal{F}_{0}}{\mathcal{F}}=2N\sqrt{\frac{%
\gamma }{\mathcal{F}}}
 \label{eqstart}
\end{equation}%
with $N=2n_{1}+n_{2}+1$ and quantum numbers $n_{1}=0,1,2,3,\ldots $, $n_{2}=0,\pm 1,\pm 2,\pm
3,\ldots ,\pm n_{1}$.  We will focus subsequently on the analysis of the ground state corresponding to $N=1$.

We solve Eq.~(\ref{eqstart}) for $\epsilon $, then expand $\epsilon $ up to
the second order in $P$ (i.e. for $v_{F}P\ll \delta $) and obtain for the exciton dispersion in Eq. (\ref{Ham_exc0})
\begin{equation}
\epsilon_{ex}(P) =-V_{0}+2\delta \sqrt{1+\frac{C}{\delta ^{3}}}+\frac{P^{2}}{2%
\mathcal{M}}\ ,  \label{eb}
\end{equation}%
where $C=\gamma (\hbar v_{F})^{2}$ and the effective exciton mass
obtained under assumption that $C\ll \epsilon (\epsilon
^{2}-4\delta ^{2})$ is given by
\begin{eqnarray}
\label{Mex}
\mathcal{M}=\frac{2\delta ^{4}}{%
v_{F}^{2}C}\sqrt{1+\frac{C}{\delta ^{3}}} \ .
\end{eqnarray}

We note the exciton effective mass $\mathcal{M}$ increase when the gap $2\delta $
increases. The exciton energy spectrum
increases with increasing gap $2\delta $.
Since annihilation of the exciton results in the emission of a
photon whose energy is that of the gap,  the renormalization
parameter $r_{0}$ must satisfy the condition
$\epsilon_{ex}(0)=2\delta$. Therefore,
\begin{equation}
r_{0}=-\frac{e^{2}}{32\pi \varepsilon _{0}\varepsilon \delta }+\frac{1}{%
4\delta }\sqrt{\frac{e^{4}}{64\pi ^{2}\varepsilon _{0}^{2}\varepsilon ^{2}}%
+2\hbar ^{2}v_{F}^{2}}\ .  \label{r0}
\end{equation}%
Then the exciton radius can be obtained from the wavefunction of the
2D harmonic oscillator and reads $\rho
=1/2\sqrt{\mathcal{F}(\epsilon (P=0))/\gamma }$. It is easy to show
that the exciton radius $\rho =1.5 r_{0}$, where we assumed that the
graphene layer was surrounded by GaAs with the dielectric constant
$\varepsilon =13$.

We note the  exciton effective mass $\mathcal{M}$ increases when the
gap parameter $\delta$ increases. The
exciton energy spectrum at the same quantum number $N$ increases
with the increase of the gap $\delta$ at small momenta $P$. The exciton energy spectrum at the same gap
$\delta$ decreases with the increase of the quantum number $N$.


\subsection{Exciton-exciton interaction}

\label{ex-ex}

Here we analyze  the Hamiltonian of exciton-exciton interaction
$\hat{H}_{ex-ex}$ in graphene in the presence of gap given by
Eq.~(\ref{Ham_exc_int}), which contributes to the exciton
Hamiltonian $\hat{H}_{ex}$ given by Eq.~(\ref{Ham_exc}).  As
discussed in Refs.~\onlinecite{Ciuti-exex} and~\onlinecite{Laikht}
for a dilute exciton gas, the excitons can be treated as bosons with
a repulsive contact interaction. For small wave vectors $q\ll \rho
^{-1}$ the pairwise exciton-exciton repulsion can be approximated as
a contact potential $U_{\mathbf{q}}\simeq U=3e^{2}\rho /(2\pi
\varepsilon _{0}\varepsilon )$. This approximation for the
exciton-exciton repulsion is
applicable, because resonantly excited excitons have very small wave vectors~%
\cite{Ciuti}. Another reason for the validity of this approximation is that
the exciton gas is assumed to be very dilute and the average distance
between excitons $r_{s}\sim (\pi n)^{-1/2}\gg \rho $, which implies the
characteristic wavenumber $q\sim r_{s}^{-1}\ll \rho ^{-1}$. A much smaller
contribution to the exciton-exciton interaction is also given by
band-filling saturation effects~\cite{Rochat}, which are neglected here.

Thus, since $U$ is directly proportional to $\rho $, and $\rho $ is
inversely proportional to $\delta $, $U$ is inversely proportional to $%
\delta $. Therefore, we can conclude that the exciton-exciton interaction $U$
decreases when $\delta $ increases.


\section{Microcavity photons}

\label{Micro photons}

The Hamiltonian of photons in a semiconductor microcavity is
determined by the cavity photon energy dispersion $\epsilon
_{ph}(P)$. According to Ref.~\onlinecite{Pau} this dispersion is
defined as $\epsilon _{ph}(P)=(c/\tilde{n})\sqrt{P^{2}+\hbar ^{2}\pi
^{2}L_{C}^{-2}},$ where $\tilde{n}=\sqrt{\varepsilon }$ is the
cavity effective refractive index that is given by the dielectric
constant of the cavity, $c$ is the speed of light in vacuum and
$L_{C}$ is the length of the microcavity.

Embedding graphene in an optical microcavity can lead to the
formation polaritons, when the excitons couple to the cavity
photons. In such a case the microcavity consists of two mirrors
parallel to each other and a graphene sheet placed in between. Then
the photons are confined in the direction perpendicular to the
mirrors, but move freely in the two directions parallel to the
mirrors. The length of the microcavity is chosen as
\begin{equation}
L_{C}=\frac{\hbar \pi c}{2\tilde{n}\delta }
\label{lc}
\end{equation}
 with the resonance
condition that the photonic and excitonic branches agree at $P=0$,
i.e. for $\epsilon_{ex}(0)=\epsilon _{ph}(0).$ This resonance
condition can be achieved either by controlling the dispersion of
excitons $\epsilon_{ex}(P)$ or by choosing the appropriate length
$L_{C}$ of the microcavity.

\section{Exciton-photon interaction}

\label{ex-ph}

We derive the Hamiltonian of the harmonic exciton-photon coupling~(\ref%
{Ham_exph}), which contributes to the total Hamiltonian of the system $\hat{H}%
_{tot}$ given by Eq.~(\ref{tot_ham}). In this Hamiltonian
exciton-photon coupling energy is represented by the Rabi splitting
constant $\hbar \Omega _{R}$. Neglecting anharmonic terms for the
exciton-photon coupling, the Rabi splitting constant $\Omega _{R}$
can be estimated quasiclassically as
\begin{equation}
\left\vert \hbar \Omega _{R}\right\vert =\left\vert \left\langle f\left\vert
\hat{H}_{int}\right\vert i\right\rangle \right\vert \ ,  \label{defrabi}
\end{equation}%
where $\hat{H}_{int}$ is the Hamiltonian of the electron-photon
interaction, $\left\vert f\right\rangle $ and $\left\vert
i\right\rangle $ are the final and initial states of the system,
correspondingly. The initial state $\left\vert i\right\rangle$
corresponds to the filled by electrons valence band and empty
conduction band, while the final state $\left\vert f\right\rangle$
corresponds to one hole in the valence band and one electron in the
conduction band. The eigenfunctions and eigenenergies of an electron
in graphene in the presence of gap are given in
Appendix~\ref{ap.wf}. For graphene this interaction is determined by
Dirac electron Hamiltonian as
\begin{eqnarray}  \label{graphint}
\hat{H}_{int}=-\frac{v_{F}e}{c}\mathbf{\sigma}\cdot
\mathbf{A}_{ph0}=\frac{v_{F}e}{i\omega }\mathbf{\sigma}\cdot
\mathbf{E}_{ph0} \  ,
\end{eqnarray}%
%
where $\mathbf{\sigma}=(%
\sigma_{x},\sigma_{y})$ are Pauli matrices, $\mathbf{A}_{ph0}$ is
the electromagnetic vector potential of a single cavity photon, and
$\ E_{ph0}=\left( 2\pi \hbar \omega /(\varepsilon W)\right) ^{1/2}\
$ is the magnitude of electric field corresponding to a single
cavity photon with frequency $\omega $ with the microcavity volume
$W$.

In Eq.~(\ref{defrabi}) the initial $|i\rangle$ and final $|f\rangle$
electron states are defined as
\begin{eqnarray}  \label{states}
| i \rangle &=& \prod_{\mathbf{q}}\hat{c}_{\mathbf{q}}^{(v)\dagger}
| 0
\rangle_{v} | 0 \rangle_{c} \ ,  \notag \\
| f \rangle &=& \hat{b}_{\mathbf{q}}^{\dagger} | i \rangle \ .
\end{eqnarray}
In Eq.~(\ref{states}), $\hat{c}_{\mathbf{q}}^{(v)\dagger}$ is the
Fermi creation operator of the electron in the valence band with the
wavevector $\mathbf{q}$, $| 0
\rangle_{c}$ denotes the wavefunction of the vacuum in the conduction band, $%
\prod_{\mathbf{q}}\hat{c}_{\mathbf{q}}^{(v)\dagger} | 0 \rangle_{v}$
corresponds to the completely filled valence band, $\hat{b}_{\mathbf{q}%
}^{\dagger}$ is the exciton creation operator with the electron in
the conduction band $c$ and the hole in the valence band $v$.
Following
Ref.~\onlinecite{Laikhtman}, $\hat{b}_{\mathbf{q}}$ and $\hat{b}_{\mathbf{q}%
}^{\dagger}$ for this case are defined as
\begin{eqnarray}  \label{opex}
\hat{b}_{\mathbf{q}} = \sum_{\mathbf{q}^{\prime}}
\hat{c}_{\mathbf{q} - \mathbf{q}^{\prime}}^{(v)\dagger}
\hat{c}_{\mathbf{q}^{\prime}}^{(c)} \ , \ \ \ \ \ \ \ \ \ \ \
\ \ \ \ \ \ \ \ \ \ \ \ \hat{b}_{\mathbf{q}}^{\dagger} = \sum_{\mathbf{q}%
^{\prime}}\hat{c}_{\mathbf{q}^{\prime}}^{(c)\dagger}
\hat{c}_{\mathbf{q} - \mathbf{q}^{\prime}}^{(v)}\ .
\end{eqnarray}
where $\hat{c}_{\mathbf{q}}^{(c)\dagger}$ and $\hat{c}_{%
\mathbf{q}}^{{c}}$ are the Fermi creation and annihilation operators
of the electron in the conduction band with the the wavevector
$\mathbf{q}$.

Substituting Eqs.~(\ref{opex}),~(\ref{efunction-c}), and ~(\ref{efunction-v}%
) into~(\ref{states}) and using the electron-photon interaction $\hat{H}%
_{int}$~(\ref{graphint}), we finally obtain from
Eq.~(\ref{defrabi}):
\begin{eqnarray}  \label{dipxy}
&& \left|\hbar \Omega_{R}\right|= \left|\frac{e v_{F}}{i \omega}
\int dx \int dy \left[\psi_{c,E_{c}}^{*}(x,y) \mathbf{\sigma}\cdot
\mathbf{E}_{ph0}
\psi_{v,E_{v}}(x,y)\right] \right|  \notag \\
&& = \left\vert \frac{e v_{F}}{2i \omega} \sqrt{\frac{\delta^{2} - E^{2}}{%
E^{2}}} \left[E_{ph0x}\left( \frac{\hbar
v_{F}(q_{x}+iq_{y})}{\delta-E} +
\frac{\hbar v_{F}(q_{x}-iq_{y})}{\delta+E} \right) \right. \right.  \notag \\
&& \left. \left. + iE_{ph0y}\left(\frac{-\hbar
v_{F}(q_{x}+iq_{y})}{\delta-E} + \frac{\hbar
v_{F}(q_{x}-iq_{y})}{\delta+E} \right) \right] \right\vert \ .
\end{eqnarray}
After simplification we obtain from Eq.~(\ref{dipxy})
\begin{eqnarray}  \label{dipxy00}
&& \left|\hbar \Omega_{R}\right|= \frac{e\hbar v_{F}^{2}}{\omega E \sqrt{%
\delta^{2} - E^{2}}} \sqrt{\left(q_{x}E_{ph0x} +
q_{y}E_{ph0y}\right)^{2}\delta^{2} + \left(q_{y}E_{ph0x} -
q_{x}E_{ph0y}\right)^{2}E^{2}} \ .
\end{eqnarray}
Assuming for simplicity electric field corresponding to a single
cavity photon mode is directed along the $x$ axis, we obtain
\begin{eqnarray}  \label{dipxy1}
\left|\hbar \Omega_{R}\right| = \frac{e v_{F}}{\sqrt{\delta^{2}+
\hbar^{2}v_{F}^{2}q^{2}}} \sqrt{\frac{2\pi\hbar}{W\varepsilon\omega}}\frac{%
\sqrt{q_{x}^{2}\delta^{2}+q_{y}^{2}E^{2}}}{q} \ .
\end{eqnarray}

In the limit $q\rightarrow 0$, when $\omega = 2 \delta/\hbar$, we
obtain from Eq.~(\ref{dipxy1})
\begin{eqnarray}  \label{dipxy2}
\left|\hbar \Omega_{R}\right| = \hbar v_{F}
e\sqrt{\frac{\pi}{W\varepsilon \delta }} \ .
\end{eqnarray}


\section{Superfluidity of microcavity polaritons}

\label{sup}

We can diagonalize the linear part of the total Hamiltonian
$\hat{H}_{tot}$ (without the second term on the right-hand side of
Eq.~(\ref{Ham_exc})) by
applying unitary transformations~\cite{Ciuti} and obtain (see Appendix~\ref%
{ap.pol}):
\begin{eqnarray}  \label{h0}
\hat{H}_{0} = \sum_{\mathbf{P}}\varepsilon_{LP}(P)\hat{l}_{\mathbf{P}%
}^{\dagger}\hat{l}_{\mathbf{P}} +\sum_{\mathbf{P}}\varepsilon_{UP}(P)\hat{u}%
_{\mathbf{P}}^{\dagger}\hat{u}_{\mathbf{P}} \ ,
\end{eqnarray}
where $\hat{l}_{\mathbf{P}}^{\dagger}$, $\hat{l}_{\mathbf{P}}$ and
$\hat{u}_{\mathbf{P}}^{\dagger}$ and $\hat{u}_{\mathbf{P}}$ are the
Bose creation and annihilation operators for the lower and upper
polaritons, respectively. The energy spectra of the low/upper
polaritons are given by Eq.~(\ref{eps0}).

Substituting the polaritonic representation of the excitonic and
photonic operators~(\ref{bog_tr}) into the total
Hamiltonian~(\ref{tot_ham}), the Hamiltonian of lower polaritons is
obtained~\cite{Ciuti}:
\begin{eqnarray}  \label{Ham_tot_p}
\hat H_{tot} = \sum_{\mathbf{P}}\varepsilon_{LP}(P)\hat{l}_{\mathbf{P}%
}^{\dagger}\hat{l}_{\mathbf{P}} + \frac{1}{2A}\sum_{\mathbf{P},\mathbf{P}%
^{\prime },\mathbf{q}} \tilde{W}_{\mathbf{P},\mathbf{P}^{\prime },\mathbf{q}%
} \hat{l}_{\mathbf{P}
+\mathbf{q}}^{\dagger}\hat{l}_{\mathbf{P}^{\prime }-
\mathbf{q}}^{\dagger}\hat{l}_{\mathbf{P}}\hat{l}_{\mathbf{P}^{\prime
}},
\end{eqnarray}
where the energy dispersion of the low polaritons
$\varepsilon_{LP}(P)$ is given by Eq.~(\ref{eps0}), and the
effective polariton-polariton interaction $\tilde{W}$ is given by
\begin{eqnarray}  \label{U_p}
\tilde{W}_{\mathbf{P},\mathbf{P}^{\prime },\mathbf{q}} = U
X_{\mathbf{P}
+\mathbf{q}} X_{\mathbf{P}^{\prime }}X_{\mathbf{P}^{\prime }-\mathbf{q}}X_{%
\mathbf{P}} \ ,
\end{eqnarray}
where $U = 3e^{2}\rho/(2\pi \varepsilon_{0} \varepsilon) $ according
to Sec.~\ref{ex-ex}.

At small momenta $\alpha \equiv 1/2 (\mathcal{M}^{-1} + (c/\tilde{n}%
)L_{C}/\hbar\pi)P^{2}/|\hbar \Omega_{R}| \ll 1$, the single-particle
lower polariton spectrum obtained from Eq.~(\ref{eps0}), in linear
order with respect to the small parameters $\alpha$, is
\begin{eqnarray}  \label{eps00}
\varepsilon_{0}(P) \approx \frac{c}{\tilde{n}} \hbar \pi L_{C}^{-1}
- |\hbar \Omega_{R}| + \frac{P^{2}}{2M_{p}} \ ,
\end{eqnarray}
where $M_{p}$ is the effective mass of polariton given by
\begin{eqnarray}  \label{Meff}
M_{p} = 2 \left(\mathcal{M}^{-1} + \frac{c L_{C}}{\tilde{n}\hbar\pi}%
\right)^{-1} \ .
\end{eqnarray}

\begin{figure}
\rotatebox{90}{
\includegraphics[width=5cm]{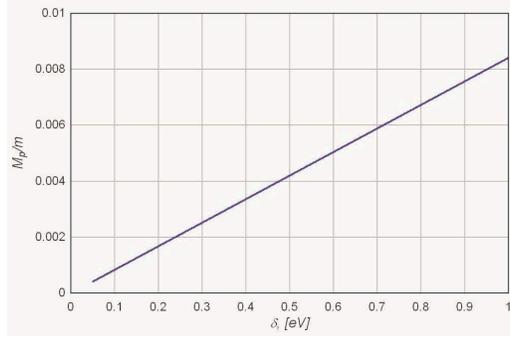}}
 \caption{The dependence of the mass of polariton $M_{p}$ on $\delta$. The polariton mass $M_{p}$ is represented in the units of electron mass $m$.}
\label{Fig_Mp}
\end{figure}

The mass of polariton $M_{p}$ as a function of  $\delta$ is
presented in Fig.~\ref{Fig_Mp}.  According to Fig.~\ref{Fig_Mp}, the
mass of polariton  $M_p$ increases when $\delta$ increases, being
directly proportional to $\delta$.

If we take into account only the lower polaritons corresponding to
the lower
energy and measure energy relative to the $P=0$ lower polariton energy $(c/%
\tilde{n}) \hbar \pi L_{C}^{-1} - |\hbar \Omega_{R}|$, the resulting
effective Hamiltonian for polaritons has the form
\begin{eqnarray}  \label{Ham_eff}
\hat H_{\mathrm{eff}} = \sum_{\mathbf{P}}\frac{P^{2}}{2M_{p}} \hat{l}_{%
\mathbf{P}}^{\dagger}\hat{l}_{\mathbf{P}} + \frac{U_{\mathrm{eff}}^{(0)}}{2A}%
\sum_{\mathbf{P},\mathbf{P}^{\prime },\mathbf{q}} \hat{l}_{\mathbf{P} +%
\mathbf{q}}^{\dagger}\hat{l}_{\mathbf{P}^{\prime }- \mathbf{q}}^{\dagger}%
\hat{l}_{\mathbf{P}}\hat{l}_{\mathbf{P}^{\prime }} \ ,
\end{eqnarray}
where $U_{\mathrm{eff}}^{(0)} = \frac{1}{4}U = 3e^{2}\rho/(8\pi
\varepsilon_{0} \varepsilon) $, since at small momenta $|X_{P}|^{2}
\approx |C_{P}|^{2} \approx 1/2$.

In the dilute limit ($n\rho^{2}\ll 1$, where $n$ is the 2D polariton
density), at zero temperature Bose-Einstein condensation (BEC) of
polaritons appears in the system, since Hamiltonian of microcavity
polaritons~(\ref{Ham_eff})  corresponds to the weakly-interacting
Bose gas. The Bogoliubov  approximation for the dilute
weakly-interacting Bose gas of polaritons results in the sound
spectrum of collective excitations at low
momenta~\cite{Abrikosov,Griffin}: $\varepsilon(P) = c_{S}P$ with the
sound velocity $c_{S} =
\left(U_{\mathrm{eff}}^{(0)}n/M_{p}\right)^{1/2} = \left(3e^{2}\rho
n/(8\pi \varepsilon_{0} \varepsilon M_{p})\right)^{1/2}$.

The dilute polaritons constructed by excitons in graphene in the
presence of the gap and microcavity photons when gapped graphene is
embedded in an optical microcavity form 2D weakly interacting gas of
bosons with the pair short-range repulsion. So the superfluid-normal
phase transition in this system is the Kosterlitz-Thouless
transition~\cite{Kosterlitz}, and the temperature of this transition
$T_c $ in a two-dimensional microcavity polariton system is
determined by the equation:
\begin{eqnarray}  \label{T_KT}
T_c = \frac{\pi \hbar ^2 n_s (T_c)}{2 k_B M_{p}}\ ,
\end{eqnarray}
where $n_s (T)$ is the superfluid density of the polariton system in
a microcavity as a function of temperature $T$, and $k_B$ is
Boltzmann constant. We obtain the superfluid density as $n_{s} = n-
n_{n}$ by determining the density of the normal component $n_{n}$
when we follow the procedure~\cite{Abrikosov} as a linear response
of the total momentum with respect to the external velocity:
\begin{eqnarray}  \label{n_n}
n_s= n- \frac{3 \zeta (3) }{2 \pi \hbar^{2}}
\frac{sk_{B}^{3}T^3}{c_S^4 M_{p}} \ ,
\end{eqnarray}
where $s=4$ is the spin degeneracy factor.

Substituting Eq.~(\ref{n_n}) for the density $n_{s}$ of the
superfluid component into Eq.~(\ref{T_KT}), we obtain an equation
for the Kosterlitz-Thouless transition temperature $T_{c}$. The
solution of this equation is
\begin{eqnarray}
\label{tct} T_c = \left[\left( 1 +
\sqrt{\frac{32}{27}\left(\frac{M_{p} k_{B}T_{c}^{0}}{\pi \hbar^{2}
n}\right)^{3} + 1} \right)^{1/3} - \left( \sqrt{\frac{32}{27}
\left(\frac{ M_{p} k_{B}T_{c}^{0}}{\pi \hbar^{2} n}\right)^{3} + 1}
- 1 \right)^{1/3}\right] \frac{T_{c}^{0}}{ 2^{1/3}} \  ,
\end{eqnarray}
where $T_{c}^{0}$ is the temperature at which the superfluid density
vanishes in the mean-field approximation (i.e., $n_{s}(T_{c}^{0}) =
0$),
\begin{equation}
\label{tct0} T_c^0 = \frac{1}{k_{B}} \left( \frac{ \pi \hbar^{2} n
c_s^4 M_{p} }{6s \zeta (3)} \right)^{1/3} \ .
\end{equation}

\begin{figure}
\rotatebox{90}{
\includegraphics[width=5cm]{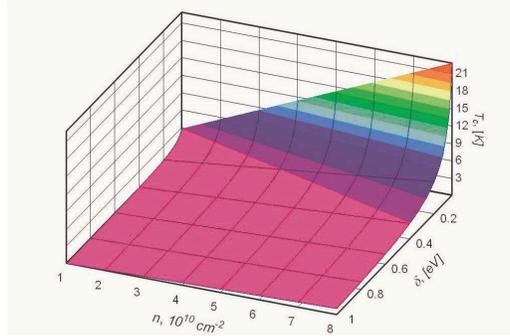}}
 \caption{The
dependence of the Kosterlitz-Thouless transition temperature $T_{c}$
on  $\delta$ and the polariton density $n$.} \label{Fig_Tc}
\end{figure}

\begin{figure}
\rotatebox{90}{
\includegraphics[width=5cm]{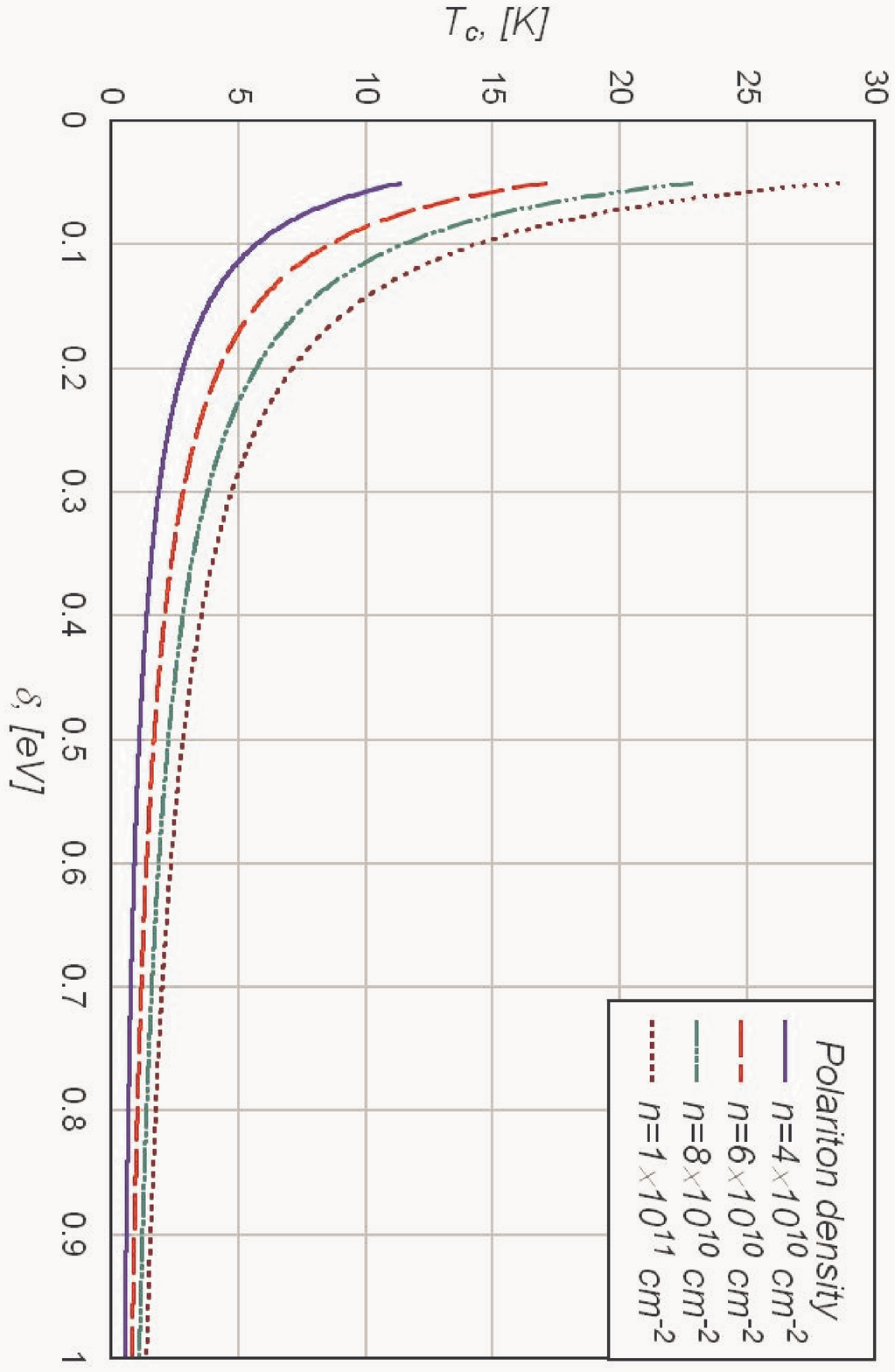}}
 \caption{The
dependence of the Kosterlitz-Thouless transition temperature $T_{c}$
on  $\delta$.} \label{Fig_Tc2D}
\end{figure}

The Kosterlitz-Thouless transition temperature $T_c$ as a function
of $\delta$ and the polariton density $n$ is presented in
Fig.~\ref{Fig_Tc}. It can be seen that $T_c$ increases,
 when  $\delta$ decreases and the polariton density $n$ increases. Fig.~\ref{Fig_Tc2D} presents the  Kosterlitz-Thouless transition temperature $T_c$ as a function
of $\delta$ at the different fixed polariton densities $n$.
According to Fig.~\ref{Fig_Tc2D}, at the same polariton density $n$,
$T_{c}$ decreases when $\delta$ increases, and at the same $\delta$,
and $T_{c}$ is higher for higher $n$.

\section{Discussion}

\label{disc}

According to Eq.~(\ref{Meff}), the effective polariton mass $M_{p}$
is mostly determined by the size of the microcavity, which depends
on the gap $\delta$ according to Eq.~(\ref{lc}) when the exciton and
microcavity photon branches of the spectrum are in resonance at zero
momentum. The effective polariton mass $M_{p}$ is proportional to
the gap $\delta$. The dependence of the superfluid density $n_s$ on
the gap $\delta$ is determined also by dependence of the sound
velocity $c_S$ on the exciton radius $\rho$ ($c_{S}\sim
\rho^{1/2}$). Since $\rho \sim \delta^{-1}$, we have $c_{S} \sim
\delta^{-1/2}$, and the normal component density $n_{n} \sim
\delta^{2}$ according to Eq.~(\ref{n_n}). The gap dependence of the
sound velocity is caused by the gap dependence of exciton-exciton
interaction described in Sec.~\ref{ex-ex}. Therefore, we conclude
that the superfluid density $n_s$ and Kosterlitz-Thouless
temperature $T_c$ are decreasing functions of the gap $\delta$ in
graphene and increasing functions of the exciton density $n$.
Besides, it follows from Eq.~(\ref{dipxy2}), that Rabi splitting is
decreasing function of the gap as $\hbar\Omega_{R} \sim
\delta^{-1/2}$.

The advantage of observing the superfluidity and BEC of polaritons
formed by gapped graphene excitons and microcavity photons in
comparison with these formed by quantum well excitons and
microcavity photons is based on the fact that the superfluidity and
BEC of polaritons formed by gapped graphene excitons can be
controlled by the gap which depends on doping. The Rabi splitting
related to the creation of an exciton in a graphene layer also can
controlled by the doping dependent gap.

In conclusion, we propose the superfluidity of 2D exciton polaritons
formed by the gapped graphene excitons and microcavity photons, when
the gapped graphene layer is embedded in an optical microcavity. The
effective exciton polariton mass is calculated as a function of the
gap energy in the graphene layers. We demonstrate the effective
exciton polariton mass increases when the gap increases and it is
directly proportional to $\delta$. We show that the superfluid
density $n_{s}$ and the Kosterlitz-Thouless temperature $T_{c}$
increases with the rise of the excitonic density $n$ and decreases
with the rise of the gap  due
to $\delta$ dependence of the sound velocity of collective excitations ($%
c_{S} \sim \delta^{-1}$), and therefore, could be controlled by $n$ and $%
\delta$. We demonstrate that the Rabi splitting related to the
creation of an exciton in a graphene layer also depends on the gap
energy.


\acknowledgments

The authors acknowledge support from the Center for Theoretical Physics of
the New York City College of Technology, CUNY and from the Deutsche
Forschungsgemeinschaft through grant ZI 305/5-1.


\appendix


\section{The eigenfunctions and eigenenergies of an electron in graphene in
the presence of gap}

\label{ap.wf}

We consider two fermions and ignore their interaction (i.e. $V(r)=0$). The
electrons in the conduction band, described by the spinor wave function $%
\psi _{cE}(x,y)$, and the holes in the valence band, described by the spinor
wave function $\psi _{vE^{\prime }}(x,y)$, are solutions of the eigenvalue
equations
\begin{equation}
H_{-\delta }\psi _{cE}=E\psi _{cE}\ ,\ \ \ H_{\delta }\psi _{vE^{\prime
}}=-E^{\prime }\psi _{vE^{\prime }}  \label{single_eigen}
\end{equation}%
of the Dirac-Weyl Hamiltonian
\begin{equation}
H_{\delta }=\left(
\begin{matrix}
\delta  & \hbar v_{F}(\partial _{x}-i\partial _{y}) \\
\hbar v_{F}(\partial _{x}+i\partial _{y}) & -\delta  \\
&
\end{matrix}%
\right) 
\ .
\end{equation}%
In the presence of a gap $2\delta $, these solutions are
\begin{eqnarray}
\label{efunction-c}
 \psi _{cE}(x,y)=\sqrt{\frac{\delta
-E}{2E}}\frac{\exp \left( i\left( q_{x}x+q_{y}y\right) \right)
}{\sqrt{L_{x}L_{y}}}\left(
\begin{array}{c}
\frac{\hbar v_{F}\left( q_{x}-iq_{y}\right) }{\delta -E} \\
1%
\end{array}%
\right) \ ,
\end{eqnarray}%
\begin{eqnarray}
\label{efunction-v} \psi _{vE}(x,y)=\sqrt{\frac{\delta
+E}{2E}}\frac{\exp \left( i\left( q_{x}x+q_{y}y\right) \right)
}{\sqrt{L_{x}L_{y}}}\left(
\begin{array}{c}
\frac{\hbar v_{F}\left( q_{x}-iq_{y}\right) }{\delta +E} \\
1%
\end{array}%
\right) \ ,
\end{eqnarray}
where $E=\sqrt{\delta ^{2}+\hbar ^{2}v_{F}^{2}q^{2}}$ is the energy of the
electron or the hole, and $L_{x}$ and $L_{y}$ are lengths in $x$ and $y$
direction, correspondingly. This allows us to construct the four components
of the spinor in Eq. (\ref{wavefunction1}) from the solutions of Eq. (\ref%
{single_eigen}) as
\begin{eqnarray}
\phi _{jk}(\mathbf{r}_{1},\mathbf{r}_{2})=\psi _{cE,j}(\mathbf{r}_{1})\psi
_{vE^{\prime },k}(\mathbf{r}_{2})\ \ \ (j=a,b;k=a,b)
\end{eqnarray}
which solves the eigenvalue equation for two non-interacting particles with
the Hamiltonian ${\mathcal{H}}$ in Eq. (\ref{ham1}) when $V(r)=0$: ${%
\mathcal{H}}_{0}\Psi =(E-E^{\prime })\Psi ,$ where ${\mathcal{H}}_{0}$ is
the Hamiltonian of two non-interacting particles.

\section{The representation of polariton operators for the Hamiltonian of
the exciton-photon system in a microcavity}

\label{ap.pol}

We express the exciton and microcavity photon operators in terms of
polariton operators. The exciton and photon operators are defined as \cite%
{Ciuti}
\begin{eqnarray}  \label{bog_tr}
\hat{b}_{\mathbf{P}} = X_{P}\hat{l}_{\mathbf{P}} -
C_{P}\hat{u}_{\mathbf{P}}
\ , \hspace{0.5cm} \hat{a}_{\mathbf{P}} = C_{P}\hat{l}_{\mathbf{P}} + X_{P}%
\hat{u}_{\mathbf{P}} \ ,
\end{eqnarray}
where $\hat{l}_{\mathbf{P}}$ and $\hat{u}_{\mathbf{P}}$ are lower
and upper polariton Bose operators, respectively, $X_{P}$ and
$C_{P}$ are
\begin{eqnarray}  \label{bog}
X_{P} = \left(1 + \left(\frac{\hbar\Omega_{R}}{\varepsilon_{LP}(P) -
\epsilon _{ph}(P)}\right)\right)^{-1/2} \ , \hspace{0.5cm} C_{P} = -
\left(1 + \left(\frac{\varepsilon_{LP}(P) - \epsilon _{ph}(P)}{%
\hbar\Omega_{R}} \right)\right)^{-1/2} \ ,
\end{eqnarray}
and the energy dispersion  of the low/upper polaritons are
\begin{eqnarray}
\varepsilon _{LP/UP}(P) &=&\frac{\epsilon _{ph}(P)+\epsilon _{ex}(P)}{2%
}  \notag  \label{eps0} \\
&\mp &\frac{1}{2}\sqrt{(\epsilon _{ph}(P)-\epsilon
_{ex}(P))^{2}+4|\hbar \Omega _{R}|^{2}}\ .
\end{eqnarray}%
%
%
We note that $|X_{P}|^{2}$ and $|C_{P}|^{2}=1-|X_{P}|^{2}$ represent
the exciton and cavity photon fractions in the lower polariton.


\end{document}